\documentclass[a4paper,11pt]{article}
\usepackage{jcappub} % for details on the use of the package, please see the JINST-author-manual
\pdfoutput=1 % if your are submitting a pdflatex (i.e., if you have
\usepackage[usenames,dvipsnames,svgnames,table]{xcolor} % use color

\usepackage[T1]{fontenc} % if needed
\usepackage[latin9]{inputenc}

\usepackage{float}
\usepackage{graphicx}
\usepackage{hyperref}
\usepackage{amsmath}
\usepackage{amssymb}
\usepackage{microtype}
\usepackage{amsbsy}
\usepackage{color}
\usepackage[capitalise]{cleveref}
\usepackage{breakurl}
\usepackage{mathtools}
\usepackage{csquotes}
\usepackage{lmodern}
\usepackage{bm}
\usepackage{dsfont}
\usepackage{bbold}
\usepackage{tensor}
\usepackage{soul}

\renewcommand*{\vec}[1]{\bm{#1}}

\newcommand*{\unit}[1]{\ensuremath{\,\mathrm{#1}}}

\newcommand{\reals}{\mathbb{R}}
\DeclareMathOperator{\diag}{diag}
\newcommand{\dderiv}{\mathrm{d}}
\newcommand{\eexp}{\mathrm{e}}

\newcommand{\hyperF}{\tensor[_2]{F}{_1}}
\newcommand{\Omegak}{\texorpdfstring{\ensuremath{\Omega_K}}{OmegaK}}
\newcommand{\order}{\mathcal{O}}

\let\phi\varphi

\newcommand{\StwoxR}{\texorpdfstring{\ensuremath{S^2 \times \reals}}{S2xR}} 
\newcommand{\HtwoxR}{\texorpdfstring{\ensuremath{\mathbb{H}^2 \times \reals}}{H2xR}}
\newcommand{\UHtwo}{\texorpdfstring{\ensuremath{\widetilde{U(\mathbb{H}^2)}}}{UH2}}

\allowdisplaybreaks

\makeatletter
\DeclareRobustCommand{\rcite}[1]{%
  \rcite@aux#1,\@nil{#1}%
}
\def\rcite@aux#1,#2\@nil#3{%
  \if\relax#2\relax
    Ref.~\cite{#3}%
  \else
    Refs.~\cite{#3}%
  \fi
}
\makeatother

\title{Cosmological constraints on anisotropic Thurston geometries } 

\author[a]{Ananda F. Smith,}
\author[a]{Craig J. Copi,}
\author[a]{and Glenn D. Starkman}

\affiliation[a]{CERCA/ISO, Department of Physics, Case Western Reserve University, \\10900 Euclid Avenue, Cleveland, OH 44106, USA}

\emailAdd{ananda.smith@case.edu}
\emailAdd{craig.copi@case.edu}
\emailAdd{glenn.starkman@case.edu}

\abstract{Much of modern cosmology relies on the Cosmological Principle, the assumption that the Universe is isotropic and homogeneous on sufficiently large scales, but it remains worthwhile to examine cosmological models that violate this principle slightly.
We examine a class of such spacetimes that maintain homogeneity but break isotropy through their underlying local spatial geometries. 
These spacetimes are endowed with one of five anisotropic model geometries of Thurston's geometrization theorem, and their evolution is sourced with perfect fluid dust and cosmological constant. 
We show that the background evolution of these spacetimes induces fluctuations in the observed cosmic microwave background (CMB) temperature with amplitudes coupled to the curvature parameter $\Omegak$. 
In order for these fluctuations to be compatible with the observed CMB angular power spectrum, we find $|\Omegak| \lesssim 10^{-5}$. 
This strongly limits the cosmological consequences of these models.}


\begin{document}
\maketitle
\flushbottom

\section{Introduction}
\label{sec:intro}
One of the main assumptions made in the current standard model of cosmology, $\Lambda$CDM, is that the Universe is spatially homogeneous and isotropic on large enough scales. 
Known as the Cosmological Principle, this assumption restricts the large-scale geometry of the Universe to one of three types, which can be presented simultaneously in a Friedmann-Lema\^{i}tre-Robertson-Walker (FLRW) metric. 
Though the Cosmological Principle is well-corroborated by large-scale structure and other cosmological observables, hints of this postulate being violated have emerged in recent decades.

Among the strongest pieces of evidence for the Cosmological Principle has been the high degree of isotropy of the blackbody temperature  of the cosmic microwave background (CMB) radiation, first reported by Penzias and Wilson \cite{PenziasWilson:1965}. 
More recently, the isotropy of the statistical properties of the very small amplitude fluctuations in the temperature and polarization of the  CMB has been tested  \cite{Abdalla:2022yfr, Planck:2019evm}, and several violations of isotropy have been reported.
In particular, a handful of large-angle features in the observed CMB temperature \cite{lowl:2004} were discovered in the first WMAP data release \cite{WMAP1:2003} and were suggested to indicate significant deviation from statistical isotropy. 
These features have come to be known as the ``large-angle anomalies,'' and their significance has persisted in  subsequent \textit{Planck} datasets \cite{Schwarz:2015cma}, with recent work arguing these anomalies jointly constitute a $ > 5\sigma$ violation of statistical isotropy \cite{Jones:2023ncn}.

This collection of evidence warrants consideration of cosmological models that deviate slightly from the Cosmological Principle, in particular by violating spatial isotropy. 
In this work, we investigate the consequences of breaking spatial isotropy through geometry -- i.e., by equipping the Universe with a background metric that is homogeneous but anisotropic. Relaxing the requirement of isotropy allows the large-scale geometry of the Universe to be of a slightly more general class than the three FLRW geometries. 

Historically, work on the cosmology of anisotropic spaces has centered around the Bianchi models (see \cite{Ellis:2006} for review). 
These are 3+1 spacetimes whose homogeneous spatial part corresponds to a 3-dimensional real Lie algebra, falling into one of eleven types within a classification devised by Bianchi in 1898 \cite{Bianchi:1898}. 
A subset of the Bianchi models have been invoked as potential explanations of CMB anomalies \cite{Rogues:2009, Sung:2009ju}, but their deviation from isotropy is strongly constrained. 
This is because many Bianchi spaces require anisotropic expansion -- i.e., multiple scale factors -- if they are sourced by perfect fluid stress energies. 
This anisotropic expansion leaves potentially detectable imprints on cosmological observables, often characterized by the resulting ``shear.''
For instance, in the subset of Bianchi spaces that explicitly contain the three FLRW metrics as special cases, anisotropic expansion is strongly constrained by the anisotropies it induces in the CMB temperature \cite{MartinezGonzalez:1995,Saadeh:2016}. 
Constraints have also been derived from the imprint of shear on nucleosynthesis \cite{BBN:1984}.

This work considers a more recently developed class of homogeneous but not necessarily isotropic geometries known as the Thurston geometries. 
These eight model geometries exhaust the possible local geometries of closed homogeneous 3-manifolds according to Thurston's geometrization theorem \cite{Thurston:1982}.
Thurston's classification schema is analogous to Bianchi's, and all but one of the Thurston geometries fall within a Bianchi group \cite{Fagundes:1985}. 
Out of the eight Thurston geometries, three are the isotropic open, closed, and flat FLRW geometries, and the remaining five are anisotropic. 
These latter subset of spaces remain largely unconstrained within a cosmological context (although see \cite{Graham:2010hh,Awwad:2022uoz}).

The dynamics and distance measures of spacetimes with Thurston geometries as spatial parts, which we will dub Thurston spacetimes, have recently been investigated \cite{Awwad:2022uoz}, albeit when equipped with a single scale factor. 
This requires invoking an anisotropic fluid fine-tuned to prohibit anisotropic expansion. 
The evolution of scale factors under isotropic dust and cosmological constant is provided for each Thurston spacetime in \cite{Awwad:2022uoz}, but no accompanying constraints on curvature scales are derived. 

In this paper, we provide strong constraints on the curvature of all five anisotropic Thurston spacetimes for the first time. 
After providing representations of each anisotropic Thurston geometry in Section \ref{sec:geometries}, we review the dynamics of the corresponding spacetimes under the presence of perfect fluid dust and cosmological constant in Section \ref{sec:evolution}, where we find anisotropic expansion is required. 
Consequently, we find that the local specific intensity of CMB photons is distorted, and a present-day observer interprets this intensity spectrum as that of a blackbody with a direction-dependent temperature. 
The amplitudes of the CMB temperature fluctuations induced in these geometries are coupled to the curvature parameter $\Omegak$, which is therefore strongly constrained by the isotropy of the CMB. 
We derive this constraint for each of the five anisotropic Thurston spacetimes in Section \ref{sec:constraints}, finding that $|\Omegak| \lesssim 10^{-5}$ in all five geometries. 

\section{Thurston geometries}
\label{sec:geometries}
In 1982 Thurston conjectured \cite{Thurston:1982} (and Perelman later proved \cite{perelman2002entropy,perelman2003ricci})
\begin{quote}
    The interior of every compact 3-manifold has a canonical decomposition into pieces which have geometric structures.
\end{quote}
In practice, this reduces to a set of eight local homogeneous three-geometries -- three being the well-known and well-studied isotropic FLRW geometries: flat ($\mathbb{R}^3$), spherical ($S^3$), and hyperbolic ($\mathbb{H}^3$).
The remaining five anisotropic local three-geometries are less well-known and will be discussed below.
Though the conjecture allows for a decomposition of the full space, in this work we restrict to the case where the observable Universe has just one of these eight local geometries. 
Without loss of generality, we take the topology to be the covering space of that geometry. 
We refer curious readers to the growing body of work on detecting nontrivial topology in the Universe, e.g., \cite{COMPACT:2022gbl}. 

In the remainder of this section, we present metrics for each of the five anisotropic Thurston geometries. 
We adopt the representations of \cite{Awwad:2022uoz}, where the parameter $\kappa$ found in the spatial part of each metric distinguishes between positive ($\kappa > 0$) and negative ($\kappa < 0$) spatial curvature. 
All of these local geometries can be made arbitrarily close to flat space in some finite neighborhood by taking $\kappa$ sufficiently close to zero.

\subsection{\boldmath \StwoxR\ and \HtwoxR}

The first two anisotropic geometries in Thurston's classification system are $\StwoxR$ and $\HtwoxR$. 
In the literature, spacetimes equipped with $\StwoxR$ spatial geometries are often referred to as Kantowski-Sachs spaces \cite{Kantowski:1966te}, and $\HtwoxR$  falls under type III of the Bianchi classification.

$\StwoxR$ and $\HtwoxR$ are the most straightforward to understand among the five geometries we are considering: $\StwoxR$ admits positive curvature ($\kappa > 0$) along two spatial directions and zero curvature along one, while $\HtwoxR$ is analogous but with its curved directions having negative curvature ($\kappa < 0$). 
This description leads naturally to a spatial metric separated into a hyperspherical part and a flat part:
\begin{equation}
    \dderiv\Sigma_3^2 = \dderiv\chi^2 + S_\kappa (\chi)^2 \dderiv\phi^2 + \dderiv z^2,
\end{equation}
where $\chi \in [0, \infty)$, $\phi \in [0, 2\pi) $, $z \in \reals$, and
\begin{equation}
    S_\kappa(\chi) = 
    \begin{cases}
        \sin(\chi \sqrt{\kappa} )/\sqrt{\kappa}, & \kappa > 0, \\
        \sinh(\chi \sqrt{-\kappa})/\sqrt{-\kappa}, & \kappa < 0.
    \end{cases}
\end{equation}

\subsection{\boldmath \UHtwo}
The next anisotropic Thurston geometry is $\UHtwo$, the universal cover of $U(\mathbb{H}^2)$, the unit tangent bundle of the hyperbolic plane, and falls within Bianchi types III and VIII. 
We adopt the spatial metric derived in \cite{Fagundes:1985},
\begin{equation}
    \dderiv\Sigma_3^2 = \dderiv x^2 + \cosh(2x\sqrt{-\kappa}) \dderiv y^2 + \dderiv z^2 + 2\sinh(x\sqrt{-\kappa})\dderiv y\dderiv z,
\end{equation}
where $\kappa < 0$ and $x, y, z \in \reals$.

This geometry is often referred to as $\widetilde{\text{SL}(2, \mathbb{R})}$ in the literature since this space is diffeomorphic to \UHtwo.
However, these two spaces are not isomorphic so they are not interchangeable in a physical context.

\subsection{Nil}

The next anisotropic Thurston geometry is Nil, the geometry of the Heisenberg group.
It can be thought of as twisted $E^2 \times \reals$ and falls within Bianchi type II.
It can be represented with the spatial metric
\begin{equation}
    \dderiv \Sigma_3^2 = \dderiv x^2 + (1+\kappa x^2) \dderiv y^2 + \dderiv z^2 - 2x\sqrt{-\kappa}\dderiv y \dderiv z,
\end{equation}
where $x, y, z \in \reals$.

\subsection{Solv}

The final anisotropic Thurston geometry is Solv, the geometry of solvable Lie groups.
The Solv geometry falls within Bianchi type VI$_0$, and can be represented with the spatial metric
\begin{equation}
    \dderiv \Sigma_3^2 = \eexp^{2z \sqrt{-\kappa}}\dderiv x^2 + \eexp^{-2z\sqrt{-\kappa}}\dderiv y^2 + \dderiv z^2,
\end{equation}
where $x, y, z \in \reals$.

\section{Evolution of anisotropic Thurston spacetimes}
\label{sec:evolution}
According to general relativity, the evolution of any spacetime is governed by the stress energy content of the universe through Einstein's field equations
\begin{equation}
    \tensor{G}{^\mu_\nu} = 8\pi G \tensor{T}{^\mu_\nu},
\end{equation}
where our choice of $c = 1$ units persists in all subsequent calculations.
Thus the evolution of spacetimes with anisotropic Thurston spatial geometries is sensitive to the choice of stress energy.
For example, it is possible to equip an anisotropic spacetime with a single scale factor, say $a(t)$, in what are commonly referred to as ``shear-free'' models in the literature \cite{Collins:1986,Mimoso:1993ym}.
The stress energy tensor characteristic of the fluid required to achieve this scales like $a^{-2}$ and contains off-diagonal elements that are chosen precisely to prevent anisotropic expansion. 
The fine-tuning required for this cancellation to occur makes this approach less compelling from a theoretical perspective and requires the introduction of non-standard sources of stress energy.

Following \cite{Awwad:2022uoz}, in this work we take the opposite approach of using known sources of isotropic stress energy and allowing anisotropic expansion.
Here our isotropic stress energy contains perfect fluids in the form of dust (pressureless with energy density $\rho$) and a cosmological constant (with $\Lambda = 8\pi G\rho_\Lambda$ and $p_\Lambda = -\rho_\Lambda$).
With this content the stress-energy tensor takes the form
\begin{equation}
    \tensor{T}{^\mu_\nu} = \rho u^\mu u_\nu - \frac{\Lambda}{8 \pi G} \tensor{\delta}{^\mu_\nu},
\end{equation}
where $u^\mu$ is the 4-velocity of the fluid. 
Working in the co-moving frame where $u^\mu = (1, 0,0,0)$ yields the diagonal stress energy tensor commonly studied in $\Lambda$CDM cosmology
\begin{equation}
    \label{eq:Tmunu}
    \tensor{T}{^\mu_\nu} = \diag\!\left(-\rho - \frac{\Lambda}{8 \pi G}, - \frac{\Lambda}{8 \pi G}, - \frac{\Lambda}{8 \pi G}, - \frac{\Lambda}{8 \pi G} \right) .
\end{equation}

In general, the spatial metric of the Thurston geometries has the form
\begin{equation}
    \dderiv\Sigma_3^2 = \gamma_{i j} \dderiv x^i \dderiv x^j,
\end{equation}
so that the spacetime metric is given by
\begin{equation}
    \dderiv s^2 = -\dderiv t^2 + \gamma_{k \ell} \tensor{\alpha}{^k_i} \tensor{\alpha}{^\ell_j} \dderiv x^i \dderiv x^j,
\end{equation}
where we allow for (potentially independent) scale factors through the diagonal matrix
\begin{equation}
    \tensor{\alpha}{^k_i} = \diag(a_1(t), a_2(t), a_3(t))\indices{^k_i}.
\end{equation}
With anisotropic expansion introduced, the Einstein field equations for the stress energy tensor \eqref{eq:Tmunu} are quite similar among the five anisotropic Thurston geometries. The diagonal elements of their Einstein field equations all have the form
\begin{alignat}{6}
    \label{eq:scalefactor-evolution}
    \indices{^0_0} &: \qquad \frac{\dot{a}_1\dot{a}_2}{a_1 a_2} &&+ \frac{\dot{a}_2\dot{a}_3}{a_2 a_3} &&+ \frac{\dot{a}_3\dot{a}_1}{a_3 a_1} &&+ \tensor{\Delta}{^0_0} && {} =~ &&\Lambda + \frac{k^{(0)}\kappa}{a_{\mathrm{dom}}^2} + 8 \pi G \rho , \nonumber \\
    \indices{^1_1} &: \qquad  \frac{\ddot{a}_2}{a_2} &&+ \frac{\ddot{a}_3}{a_3} &&+ \frac{\dot{a}_2\dot{a}_3}{a_2 a_3} &&+ \tensor{\Delta}{^1_1} && {} =~ &&\Lambda + \frac{k^{(1)}\kappa}{a_{\mathrm{dom}}^2}, \\
    \indices{^2_2} &: \qquad  \frac{\ddot{a}_3}{a_3} &&+ \frac{\ddot{a}_1}{a_1} &&+ \frac{\dot{a}_3\dot{a}_1}{a_3 a_1} &&+ \tensor{\Delta}{^2_2} && {} =~ &&\Lambda + \frac{k^{(2)}\kappa}{a_{\mathrm{dom}}^2}, \nonumber \\
    \indices{^3_3} &: \qquad  \frac{\ddot{a}_1}{a_1} &&+ \frac{\ddot{a_2}}{a_2} &&+ \frac{\dot{a}_1\dot{a}_2}{a_1 a_2} &&+ \tensor{\Delta}{^3_3} && {} =~ &&\Lambda + \frac{k^{(3)}\kappa}{a_{\mathrm{dom}}^2}. \nonumber
\end{alignat}
The dependence of \eqref{eq:scalefactor-evolution} on the choice of geometry lies in the term proportional to $\kappa$, where the constants $k^{(\mu)}$ and accompanying scale factor $a_{\mathrm{dom}}$ are listed in Table \ref{tab:params}, and in extra contributions $\tensor{\Delta}{^\mu_\nu}$ to the Einstein tensor.
We define $\tensor{\Delta}{^\mu_\nu}$ to include all of the off-diagonal elements of the Einstein tensor, meaning the accompanying off-diagonal field equations are, given our insistence on a diagonal stress-energy tensor, simply
\begin{equation}\label{eq:off-diag}
    \tensor{\Delta}{^\mu_\nu} = 0.
\end{equation}

\begin{table}[b]
    \centering
    \begin{tabular}{|c|cccc|c|}
    \hline
        Geometry & $k^{(0)}$ & $k^{(1)}$ & $k^{(2)}$ & $k^{(3)}$ & $a_{\mathrm{dom}}$ \\
        \hline
        $S^2 \times \mathbb{R}$ \text{ and } $\mathbb{H}^2 \times \mathbb{R}$ & -1 & 0 & 0 & -1 & $a_1\equiv a$\\
        $\widetilde{U(\mathbb{H}^2)}$ & -5/4 & 1/4 & 1/4 & -7/4 & $a_1\equiv a$ \\
        Nil & -1/4 & 1/4 & 1/4 & -3/4 & $a_1\equiv a$\\
        Solv & -1 & -1 & -1 & 1 & $a_3\equiv b$ \\
        \hline
    \end{tabular}
    \caption{Curvature coupling constants $k^{(i)}$ and the scale factor in the curvature term $a_{\mathrm{dom}}$ for each of the five anisotropic Thurston spacetimes.}
    \label{tab:params}
\end{table}

The evolution of the scale factors are constrained by \eqref{eq:off-diag}, since the off-diagonal elements of $\tensor{\Delta}{^\mu_\nu}$ will not necessarily be zero if all of the scale factors evolve independently.
We will show that the solutions to the diagonal field equations \eqref{eq:scalefactor-evolution} equate certain pairs of scale factors in a way that satisfies \eqref{eq:off-diag} in the limit that the spatial curvature is small, which is mandated by the Universe being approximately flat on large scales.
To this end, we may define an average scale factor as the geometric mean
\begin{equation}
    \label{eq:scale-factor-t}
    A(t) \equiv \left[ a_1(t) a_2(t) a_3(t) \right]^{1/3},
\end{equation}
the Hubble parameter associated with it
\begin{equation}
    \label{eq:Hmean}
    H(t) \equiv \frac{\dot{A}}{A} = \frac{1}{3} \left( \frac{\dot{a}_1}{a_1} + \frac{\dot{a}_2}{a_2} + \frac{\dot{a}_3}{a_3} \right),
\end{equation}
and the dimensionless curvature fraction
\begin{equation}
    \label{eq:OmegaK}
    \Omegak \equiv \frac{k^{(0)} \kappa}{3 H_0^2},
\end{equation}
where $H_0\equiv H(t_0)$ is the current ($t=t_0$) value of the Hubble expansion parameter.
More precisely, then, the small curvature limit corresponds to $|\Omegak|\ll 1$, ensuring the terms proportional to $\kappa$ on the right-hand side of \eqref{eq:scalefactor-evolution} are subdominant.

With this, taking linear combinations of the spatial equations in \cref{eq:scalefactor-evolution}, and following \cite{Awwad:2022uoz} we can expand the individual scale factors in powers $\Omegak$ to find
\begin{equation}
    \label{eq:scalefactors-t}
    a_i(t) = A(t) \left[ 1 + \Omegak K^{(i)} F(t) + \order(\Omegak^2) \right],
\end{equation}
where 
\begin{align}
    \label{eq:Ks}
    K^{(1)} &\equiv \frac{k^{(2)} + k^{(3)} - 2 k^{(1)}}{k^{(0)}} , \nonumber \\
    K^{(2)} &\equiv \frac{k^{(3)} + k^{(1)} - 2 k^{(2)}}{k^{(0)}} , \\
    K^{(3)} &\equiv \frac{k^{(1)} + k^{(2)} - 2 k^{(3)}}{k^{(0)}}  \nonumber
\end{align}
are given in \cref{tab:Ks} and
\begin{equation}\label{eq:F}
    F[A(t)] \equiv \frac{2}{5\Omega_m} \int_{A(t_0)}^{A(t)} \, \frac{\hyperF\!\left(\frac{1}{2}, \frac{5}{6}; \frac{11}{6}; -a'^3 \frac{\Omega_\Lambda}{\Omega_m}\right)}{\sqrt{1 + a'^3 \frac{\Omega_\Lambda}{\Omega_m}}} \, \dderiv a' \,.
\end{equation}
Here $\hyperF$ is the hypergeometric function, the integral is from the time  $t_0$, and we have made the standard definitions
\begin{equation} 
    \Omega_m \equiv \frac{8\pi G \rho(t_0)}{3 H_0^2}\, \mbox{ and } \Omega_\Lambda \equiv \frac{\Lambda}{3 H_0^2}.
\end{equation}

\begin{table}[b]
    \centering
    \begin{tabular}{|c|ccc|}
        \hline
        Geometry & $K^{(1)}$  & $K^{(2)}$ & $K^{(3)}$\\
        \hline
        $\StwoxR$ \text{ and } $\HtwoxR$ & $\hphantom{-}1$ & $\hphantom{-}1$ & $-2$ \\
        $\UHtwo$ & $\hphantom{-} 8/5$ & $\hphantom{-} 8/5$ & $-16/5$ \\
        Nil & $\hphantom{-} 4$ & $\hphantom{-} 4$ & $-8$ \\
        Solv & $-2$ & $-2$ & $\hphantom{-} 4$ \\
        \hline
    \end{tabular}
    \caption{Curvature coupling differences $K^{(i)}$ as defined in \eqref{eq:Ks} for the five anisotropic Thurston spacetimes.}
    \label{tab:Ks}
\end{table}

While these same solutions were obtained in \cite{Awwad:2022uoz} and we have followed a similar procedure, there is an important difference.
Namely, in that work, $\tensor{\Delta}{^\mu_\nu}$ was set to zero by equating scale factors as necessary in each geometry \emph{before} solving the field equations.  
In $\widetilde{U(\mathbb{H}^2)}$ and Nil, this meant imposing the restriction $a_1 = a_2 = a_3$ and $a_2 = a_3$, respectively.
However, this is inconsistent with \eqref{eq:scalefactors-t} given the values of $K^{(i)}$ in \cref{tab:Ks} which indicates that $a_1 = a_2 \neq a_3$ to order $\Omegak$ in all geometries.
To avoid this contradiction we instead use \eqref{eq:scalefactors-t} along with the $K^{(i)}$ from \cref{tab:Ks} to determine which scale factors to equate.
With these solutions, we find that $\tensor{\Delta}{^\mu _\nu}$  still vanishes to order $\Omegak$ across all geometries. 
In \StwoxR, \HtwoxR, and Solv, this is because the only non-trivial entries of $\tensor{\Delta}{^\mu _\nu}$ exhibit the proportionality
\begin{equation}
    \tensor{\Delta}{^0_1} = -a_1^2 \tensor{\Delta}{^1_0} \propto \left(\frac{\dot{a_1}}{a_1} - \frac{\dot{a_2}}{a_2}\right),
\end{equation}
which vanishes by the equivalence of $a_1$ and $a_2$ to this order.
In the remaining two geometries, $\UHtwo$ and Nil, the entries of $\tensor{\Delta}{^\mu_\nu}$ are less obviously zero to working order and take the form
\begin{equation}\label{eq:Uh2_delta}
    \tensor{\Delta}{^\mu _\nu} = \sqrt{\Omegak}f(a,b)\,.
\end{equation}
Here $f(a,b)$ contains differences of scale factors $a_1 = a_2 \equiv a$ and $a_3 \equiv b$.
Thus $f$ must be at least $\order(\sqrt{\Omegak})$ since it must disappear in the flat limit, i.e., when $\Omegak \to 0$.
In practice $f$ is found to be $\order(\Omegak)$ as shown in Appendix \ref{app:delta_entries}, where more detailed expressions for \eqref{eq:Uh2_delta} are listed. 
We proceed knowing that we may neglect $\tensor{\Delta}{^\mu _\nu}$ to working order in $\Omegak$.

Turning to the evolution of the average scale factor $A(t)$, we may square \cref{eq:Hmean} to find
\begin{equation}\label{eq:H-squared}
    H^2
    = \frac{1}{9} \sum_{i=1}^3 \left( \frac{\dot{a}_i}{a_i} \right)^2 + \frac{2}{9} \left(\Lambda + \frac{k^{(0)}\kappa}{a_{\mathrm{dom}}^2} + 8\pi G\rho \right),
\end{equation}
where we have used $\tensor{\Delta}{^\mu_\nu}=0$ to the required order and have used the $^0\text{}_0$ equation from \eqref{eq:scalefactor-evolution} to simplify the second term.
It follows from \eqref{eq:scalefactors-t} that
\begin{equation}\label{eq:hubble-rates-sum}
    \sum_{i=1}^3 \left( \frac{\dot{a}_i}{a_i} \right)^2 
    = 3 H^2 + \Omegak q(t) \sum_i K^{(i)} + \order(\Omegak^2).
\end{equation}
for a known function $q(t)$.
The form of $q(t)$ is unimportant since from \eqref{eq:Ks} 
\begin{equation}
    \sum_{i=1}^3 K^{(i)} = 0\,,
\end{equation}
independent of geometry.
Combining \eqref{eq:H-squared} and \eqref{eq:hubble-rates-sum}, we obtain the Friedmann equation
\begin{equation}
    \label{eq:FriedmaneqnforA}
    H^2 = \left( \frac{\dot{A}}{A} \right)^2 = H_0^2 \left( \frac{\Omega_m}{A^3} + \Omega_\Lambda + \frac{\Omegak}{A^2} \right) + \order(\Omegak^2).
\end{equation}
Expanding in powers of $\Omegak$
\begin{equation}
    A(t) = A^{(0)}(t) + \Omegak A^{(1)}(t) + \order(\Omegak^2),
\end{equation}
and substituting into \eqref{eq:FriedmaneqnforA} we find the desired evolution.
As expected, the zeroth order term is the FLRW expansion factor for a matter and cosmological-constant-dominated universe
\begin{equation}
    \label{eq:meanscalefactor-t}
    A^{(0)}(t) = \left( \frac{\Omega_m}{\Omega_\Lambda} \right)^{1/3} \sinh^{2/3} \!\left(\frac{3\sqrt{\Omega_\Lambda}}{2} H_0 t \right).
\end{equation}
We can also obtain an explicit form for  $A^{(1)}(t)$, however, it will not be needed in subsequent calculations.
Finally, combining these results with \cref{eq:scalefactors-t} we arrive at the fully expanded time evolution for the scale factors
\begin{equation}
    \label{eq:scalefactor-expansion}
    a_i(t) = A^{(0)}(t) + \Omegak \left[ A^{(1)}(t) + K^{(i)} A^{(0)}(t) F(A^{(0)}(t)) \right] + \order(\Omegak^2).
\end{equation}

We have thus demonstrated that homogeneous spacetimes with anisotropic Thurston geometries are able to evolve under a perfect fluid stress energy to  $\mathcal{O}(\Omegak)$ when multiple scale factors are introduced.
This expansion induces angular dependence in cosmological observables and distance measures.
In particular, a completely isotropic CMB at last scattering will be observed to be anisotropic by a post-last-scattering observer, such as ourselves.
In the following section, we show that these fluctuations allow us to put strong bounds on $\Omegak$ should our Universe possess these anisotropic local geometries.

\section{\boldmath Constraints from induced CMB temperature anisotropies}
\label{sec:constraints}
It has been well-known since Penzias and Wilson's initial detection \cite{PenziasWilson:1965} that the temperature of the CMB is nearly uniform across the sky.
Microkelvin fluctuations about this nearly uniform temperature have also been measured to high precision in all-sky surveys of the microwave sky and, in $\Lambda$CDM cosmology, are commonly attributed to density perturbations present at the time of recombination.
In spacetimes with anisotropic Thurston geometries as their spatial parts, there is an additional source of CMB temperature anisotropies independent of primordial density perturbations.
Namely, the geodesic trajectories of CMB photons propagating from recombination onwards will be warped in a direction-dependent way by the underlying spatial anisotropy.
This will induce angular dependence in the specific intensity of CMB photons measured by a post-last-scattering local observer,
\begin{equation}
    I(E) = \frac{E \,\dderiv N\!(E)}{\dderiv \Omega \, \dderiv A \dderiv t \, \dderiv E},
\end{equation}
where $\dderiv N(E)$ is the number of photons with energies between $E$ and $E + \dderiv E$ received within a solid angle $\dderiv \Omega = \dderiv (\cos \theta) \dderiv \phi$ in an area $\dderiv A$ during a time $\dderiv t$. 
This observer will interpret the directional-dependence in the measured specific intensity as the CMB possessing a direction-dependent temperature.

To compute these induced temperature anisotropies, we may first take the specific intensity of the CMB at recombination to be that of a blackbody with a uniform temperature $T_r$
\begin{equation}\label{eq:I_r}
    I_r(E_r) = \frac{E_r \, \dderiv N_r(E_r)}{\dderiv \Omega_r \, \dderiv A_r \dderiv t_r \, \dderiv E_r} = \frac{p E_r^3}{\eexp^{E_r/T_r}-1},
\end{equation}
where $p = 1/(2\pi)^2$ in natural units ($\hbar = c = k = 1$).
We may obtain the specific intensity measured by a present-day observer by performing a change of coordinates from quantities at recombination ($\Omega_r$, $A_r$, $t_r$, $E_r$) to their present-day counterparts ($\Omega_0$, $A_0$, $t_0$, $E_0$).
The effect of this change on the specific intensity is captured through the appropriate Jacobian factor,
\begin{equation}\label{eq:I_transformed}
    I_0(E_0) = I_r(E_r) \left(\frac{\dderiv \Omega_r}{\dderiv \Omega_0}\right) \left(\frac{\dderiv A_r \dderiv t_r}{\dderiv A_0 \dderiv t_0}\right) \left(\frac{E_0 \dderiv E_r}{E_r \dderiv E_0} \right).
\end{equation}
Knowing \eqref{eq:I_r}, we know the functional form of this present-day specific intensity to be
\begin{equation}\label{eq:I_0}
    I_0(E_0) = \frac{p E_0^3}{\eexp^{E_0/T}-1} \left(\frac{\dderiv \Omega_r}{\dderiv \Omega_0}\right) \left(\frac{\dderiv A_r \dderiv t_r}{\dderiv A_0 \dderiv t_0}\right) \left(\frac{E_r^2 \dderiv E_r}{E_0^2 \dderiv E_0} \right),
\end{equation}
where we define the present-day temperature to be
\begin{equation}\label{eq:temp_today}
    T \equiv T_r \left(\frac{E_r}{E_0}\right).
\end{equation}

In principle, angular dependence could be contained in the present-day specific intensity \eqref{eq:I_0} within the temperature \eqref{eq:temp_today} or within the Jacobian factor (i.e., a greybody factor). From Liouville's theorem, ignoring any scattering, the position-momentum phase-space volume $\dderiv V\!(E)$ occupied by $\dderiv N\!(E)$ photons of a specified energy $E$ is conserved along photon geodesics \cite{Lindquist:1966igj}: there is no greybody factor.
As a corollary, the phase-space number density $n(E) \equiv \dderiv N(E)/\dderiv V(E)$ of photons must also be conserved, and it can be shown that
\begin{equation}
    n(E) = \frac{I(E)}{E^3}
\end{equation}
in natural units \cite{Misner:1973prb}.
By demanding $n$ be the same today as at recombination, the specific intensity measured today is straightforwardly 
\begin{equation}
    I_0(E_0) = I_r(E_r) \left(\frac{E_0}{E_r}\right)^3 = \frac{p E_0^3}{e^{E_0/T}-1}
\end{equation}
for $T$ given by \eqref{eq:temp_today}.
In other words, Liouville's theorem ensures that the Jacobian factor of \eqref{eq:I_0} equals unity; a collection of photons with a blackbody spectrum retains that blackbody spectrum as it propagate collectively without scattering along a null geodesic.
% \GDS{Note, I think  that this would not be true if these were different null geodesics (with, in principle, different evolutions of $E^0$), so it is important that it is a single null geodesic.}
% \GDS{(We are neglecting the spectral distortions of the blackbody that may already exist at recombination.)}
% \CJC{Is this relevant? If there were spectral distortions would they not just get carried along? Maybe we instead want to say that a blackbody remains a blackbody without directly calling it ``the CMB''.}

The task of computing the CMB temperature fluctuations induced in anisotropic spaces is then reduced to simply computing \eqref{eq:temp_today}, the ratio between photon energies measured today and at recombination, or, equivalently, the redshift of CMB photons
\begin{equation}\label{eq:redshift}
    1 + z = \frac{E_r}{E_0}.
\end{equation}
One way to compute this redshift is to solve the geodesic equations in each geometry, define the angle $\theta$ as the angle of a photon's trajectory relative to the $z$-axis, and relate how $\theta$ changes between recombination and today to the redshift of CMB photons.
Alternatively it is expedient to instead frame the problem as a Sachs-Wolfe effect \cite{SachsWolfe}, which describes how small amplitudes perturbations atop a flat FLRW metric induce temperature anisotropies in the CMB.
More precisely, we may write the spacetime metric $\tensor{g}{_\mu_\nu}$ in the form
\begin{equation}\label{eq:flat_and_h}
    \tensor{g}{_\mu _\nu} = \left[ A^{(0)}(\eta) \right]^2[\tensor{\bar{g}}{_\mu _\nu} + \tensor{h}{_\mu _\nu}],
\end{equation}
where the co-moving time $t$ has been exchanged for the conformal time,
\begin{equation}
    \eta = \int A^{(0)}(t) dt,
\end{equation}
making the flat FLRW metric
$\tensor{\bar{g}}{_\mu _\nu} = \diag(-1, 1, 1, 1)$.
It was shown in \cite{SachsWolfe} that the redshift of CMB photons in spacetimes with metrics of the form \eqref{eq:flat_and_h} is reduced to an integral along the geodesics of the unperturbed spacetime,
\begin{equation}\label{eq:1+z}
    1 + z = \frac{A^{(0)}(\eta_r)}{A^{(0)}(\eta_0)}\left[1 - \frac{1}{2}\int_{\eta_r}^{\eta_0}\left(\frac{\partial \tensor{h}{_i _j}}{\partial \eta}e^i e^j - 2 \frac{\partial \tensor{h}{_j_0}}{\partial \eta}e^j \right)\dderiv \lambda \right],
\end{equation}
to linear order in $\tensor{h}{_\mu _\nu}$, where indices run over spatial coordinates, $e^i \equiv (\sin \theta \cos \phi, \sin \theta \sin \phi, \cos \theta)$, and $\eta_r$ and $\eta_0$ are the conformal times at recombination and today, respectively.

We know that $\tensor{h}{_j _0} = \tensor{h}{_0 _j} = 0$ in spacetimes with one of the five anisotropic Thurston geometries, because we are only modifying the spatial part of the spacetime metric.
Furthermore, in these spacetimes we may write the metric perturbation to $\order(\Omegak)$ as a background flat FLRW-metric plus a correction to the spatial components of the metric:
\begin{equation}
    \tensor{h}{_i _j} = 
    \begin{cases}
        \sqrt{\Omegak} \alpha_i(\vec{x}) + 2\Omegak\left[K^{(i)}F(\eta) + \frac{A^{(1)}(\eta)}{A^{(0)}(\eta)} + \beta_i(\vec{x}) \right], & i = j, \\
        \sqrt{\Omegak}\zeta_{ij} (\vec{x}), & i \neq j,
    \end{cases}
\end{equation}
where $\alpha_i(\vec{x})$, $\beta_{i}(\vec{x})$, and $\zeta_{ij}(\vec{x})$ are geometry-dependent functions that depend only on spatial coordinates $\vec{x}$; they do not depend on $\Omegak$ nor on time.
The partial derivatives with respect to $\eta$ in \eqref{eq:1+z} ensure that the only components of $\tensor{h}{_i _j}$ that contribute to redshift are the components that depend explicitly on time, so that \eqref{eq:1+z} reduces to
\begin{equation}\label{eq:1+z_2}
    1 + z \approx \frac{A^{(0)}(\eta_r)}{A^{(0)}(\eta_0)}\left[1 - \Omegak \int_{\eta_r}^{\eta_0} \frac{\partial}{\partial \eta} \! \left(K^{(i)}F(\eta) + \frac{A^{(1)}(\eta)}{A^{(0)}(\eta)}\right)e^i e^i \dderiv \lambda \right],
\end{equation}
where $\approx$ denotes equivalence to $\order(\Omegak)$.
Knowing $K^{(1)} = K^{(2)} = -K^{(3)}/2$ in each geometry and that $e^i$ stays constant along the unperturbed (radial) photon geodesics, we may evaluate the integral \eqref{eq:1+z_2} explicitly, yielding
\begin{equation}
    1 + z \approx \frac{A^{(0)}(\eta_r)}{A^{(0)}(\eta_0)}\left. \left[1 - \frac{1}{2}\Omegak K^{(3)}\left(3 \cos^2 \theta  - 1\right) F(\eta) - \Omegak \frac{A^{(1)}(\eta)}{A^{(0)}(\eta)} \right]\right|_{\eta_r}^{\eta_0}.
\end{equation}

Since we are interested in temperature anisotropies, it will be useful to define $T_0$ as the monopole of the present-day CMB temperature, given by
\begin{equation}
    T_0 \approx T_r\left(\frac{A^{(0)}(\eta_0)}{A^{(0)}(\eta_r)}\right)\left. \left(1+\Omegak \frac{A^{(1)}(\eta)}{A^{(0)}(\eta)}\right)\right|_{\eta_r}^{\eta_0}.
\end{equation}
It follows from \eqref{eq:temp_today} that the present-day CMB temperature in the direction $(\theta,\phi)$ is 
\begin{equation}
    T(\theta,\phi) \approx T_0\left[1-\frac{1}{2}\Omegak K^{(3)}(3 \cos^2\theta - 1)F(\eta_r)\right] + \order (\Omegak^{3/2})\,,
\end{equation}
since $F(\eta_0)$ is zero by definition in \eqref{eq:F}.
The corresponding temperature anisotropies in each geometry are then
\begin{equation}
    \Delta T \equiv T - T_0 \approx -\frac{1}{2}\Omegak K^{(3)}(3\cos^2 \theta - 1) T_0 F(\eta_r) = -2\sqrt{\frac{\pi}{5}} K^{(3)}F(\eta_r)T_0\Omegak Y_{20}(\theta, \phi).
\end{equation}
A pure $Y_{20}$ harmonic with an amplitude proportional to $\Omegak$ is induced in each of the five geometries.
The power  induced in the $\ell = 2$ mode is
\begin{equation}
    D_{2} \approx \frac{12}{25} (K^{(3)})^2 F(\eta_r)^2 T_0^2 \Omegak^2,
\end{equation}
where we have used
\begin{equation}
    D_\ell = \frac{\ell(\ell+1)}{2\pi} C_\ell, \qquad C_\ell = \frac{1}{2\ell + 1}\sum_{m = -\ell}^{\ell}|a_{\ell m}|^2.
\end{equation}
By requiring that the power in this induced quadrupole must not exceed the power in the observed quadrupole \cite{Planck:2019nip},
\begin{equation}
    D_2^{\mathrm{obs}} = 225.9 \unit{\mu K^2}
\end{equation}
we find that $\Omegak$ is bounded by\footnote{Note that there is no need to account for cosmic variance on this induced quadrupole, since it is a deterministic consequence of the mean temperature at recombination in this geometry.}
\begin{equation}\label{eq:Omegak_bound}
    |\Omegak| \lesssim \frac{\left(\frac{25}{12}D_2^{\mathrm{obs}}\right)^{1/2}}{|K^{(3)}F(\eta_r)|T_0}.
\end{equation}

Using the known mean CMB temperature $T_0 = 2.726$ K \cite{Mather:1993ij}, $F(\eta_r) \approx -1.04$ from direct integration of \eqref{eq:F}, and $K^{(3)} \approx \order(1)$ in all geometries, \eqref{eq:Omegak_bound} requires $|\Omegak| \lesssim 10^{-5}$. 
This is considerably more stringent than the $\Omegak = 0.001 \pm 0.002$ bound reported by \textit{Planck} \cite{Planck:2018vyg} for the isotropic FLRW spaces.

% \section{\boldmath \UHtwo\ and Nil}
% \label{sec:uh2 and nil}
% \input{text/uh2_and_nil}

% \section{Solv}
% \label{sec:solv}
% \input{text/solv}

\section{Conclusion}
\label{sec:conclusion}
In this paper, we have, for the first time, derived powerful constraints on  $\Omegak$ in spacetimes with homogeneous anisotropic spatial geometries of all five corresponding Thurston types -- $S^2 \times \reals$, $\mathbb{H}^2 \times \reals$, $\widetilde{U(\mathbb{H}^2)}$, Nil, and Solv -- when filled with standard homogeneous  perfect-fluid dust and cosmological constant $\Lambda$.
We have shown that, in all geometries, CMB photons undergo direction-dependent redshift as they propagate from recombination through anisotropically expanding space. 
An observer interprets these effects as a blackbody with temperature possessing a quadrupolar anisotropy with an amplitude proportional to a power of  $\Omegak$. 
We find that, in order for the induced quadrupole to be no larger than the observed quadrupolar fluctuations in the CMB temperature, $|\Omegak| \lesssim 10^{-5}$ in all five geometries. 

We have not commented on the impact of primordial perturbations on the observed CMB temperature. 
Though it is theoretically possible for this anisotropy to be exactly canceled out by temperature fluctuations existing at the time of last scattering, it is highly unlikely that we live just at the moment when such cancellation occurs.

We emphasize that these constraints are valid for a Universe filled with standard homogeneous  perfect-fluid dust and cosmological constant, but do not necessarily hold for less conventional stress-energy contents. 
For instance, it has been shown that the CMB temperature maintains its isotropy in certain shear-free models sourced by a finely-tuned anisotropic fluid \cite{PhysRevD.83.023509}. 
We note however that the angular diameter and luminosity distances are still direction-dependent in the shear-free renditions of anisotropic Thurston spacetimes \cite{Awwad:2022uoz}, so it still may be possible to constrain anisotropy less stringently using other cosmological observables (e.g., see \cite{Pereira:2015pxa}).

Our powerful constraints come from comparison of observed CMB temperature fluctuations with 
the direction-dependent specific intensity that results from the evolution of an isotropic homogeneous blackbody distribution at the time of recombination into an anisotropic photon distribution today.
These constraints are therefore independent of the physics responsible for the usual primordial fluctuations.

\acknowledgments
We thank Yashar Akrami for early conversations identifying the questions addressed in this paper, and thank Johanna Nagy and John Ruhl for useful conversations about CMB measurements.
A.F.S. was partially supported by NASA ATP Grant RES240737;
G.D.S. by DOE grant DESC0009946.

\appendix

\section{\texorpdfstring{$\tensor{\Delta}{^\mu_\nu}$}{Δ mu nu} in \UHtwo\ and Nil}

\label{app:delta_entries}
Here we list out the entries of $\tensor{\Delta}{^\mu _\nu}$ in $\UHtwo$ and Nil.
For $\mu, \nu \in \{0,1\}$ 
\begin{alignat}{3}
    \label{eq:Delta00}\tensor{\Delta}{^0_0} & =~ -\tensor{\Delta}{^1_1} & =& -x^2\frac{3\Omegak}{4k^{(0)}}\left[\frac{\dot{a}}{a} - \frac{\dot{b}}{b} \right]^2 + \order(\Omegak^2) \\
    \label{eq:Delta01}\tensor{\Delta}{^0_1} & = -a^2\tensor{\Delta}{^1_0} & = & - x \frac{3\Omegak}{2 k^{(0)}}\left[\frac{\dot{a}}{a} -\frac{\dot{b}}{b} \right] + \order(\Omegak^2),
\end{alignat}
where we have set $a_1 = a_2 = a$ and $a_3 = b$. 

The remaining nontrivial elements of $\tensor{\Delta}{^\mu _\nu}$ are 
\begin{alignat}{3}
    \label{eq:Delta22}\tensor{\Delta}{^2_2} & =~~ & -x^2\frac{3 \Omegak}{4k^{(0)}} \left[\left(\frac{\dot{a}}{a}\right)^2 -\left(\frac{\dot{b}}{b}\right)^2 + \frac{2\ddot{a}}{a} - \frac{2\ddot{b}}{b} \right]& + \order(\Omegak^2) \\
    \tensor{\Delta}{^2_3} & =~~ & \zeta x\sqrt{-\frac{3\Omegak}{4k^{(0)}}}\left(\frac{b}{a} \right) \left[ - \frac{\dot{a}\dot{b}}{ab} + \left(\frac{\dot{b}}{b}\right)^2 - \frac{\ddot{a}}{a} + \frac{\ddot{b}}{b}  \right]& + \order(\Omegak^{3/2})\\
    \tensor{\Delta}{^3_2} & =~~ & \zeta x\sqrt{-\frac{3\Omegak}{4k^{(0)}}}\left(\frac{a}{b} \right) \left[2\left(\frac{\dot{a}}{a}\right)^2 - \frac{\dot{3a}\dot{b}}{ab} + \left(\frac{\dot{b}}{b}\right)^2 + \frac{\ddot{a}}{a} - \frac{\ddot{b}}{b}  \right]& + \order(\Omegak^{3/2})\\
    \label{eq:Delta33}\tensor{\Delta}{^3_3} & =~~ & x^2\frac{3 \Omegak}{4k^{(0)}} \left[3\left(\frac{\dot{a}}{a}\right)^2 -\frac{4\dot{a}\dot{b}}{ab}+\left(\frac{\dot{b}}{b}\right)^2 + \frac{2\ddot{a}}{a} - \frac{2\ddot{b}}{b} \right]& + \order(\Omegak^2),
\end{alignat}
where $\zeta = +1$ in Nil and $\zeta = -1$ in $\UHtwo$.

From \eqref{eq:scalefactors-t}, we know $a$ and $b$ differ by an order $\Omegak$ correction, meaning \eqref{eq:Delta00} and \eqref{eq:Delta01} are high enough order in $\Omegak$ to be neglected.
We may verify the same is true for the other nontrivial elements  by using \eqref{eq:scalefactors-t} to rewrite \eqref{eq:Delta22}--\eqref{eq:Delta33} as
\begin{alignat}{5}
    \tensor{\Delta}{^2_2} & =~ & -\tensor{\Delta}{^3_3} & =~ & -x^2\frac{3\Omegak^2}{2k^{0}}(K^{(1)} - K^{(3)})\left(\frac{3\dot{A}\dot{F}}{A} + \ddot{F} \right) + \order(\Omegak^2)\\
    \tensor{\Delta}{^2_3} & =~ & -\tensor{\Delta}{^3_2} & =~ & -\zeta x\sqrt{\frac{-3\Omegak^3}{4k^{0}}}(K^{(1)} - K^{(3)})\left(\frac{3\dot{A}\dot{F}}{A} + \ddot{F} \right) + \order(\Omegak^{3/2}),
\end{alignat}
where $F = F[A(t)]$ is given by \eqref{eq:F}.

% \section{Temperature fluctuation amplitudes in \UHtwo\ and Nil}
% \input{text/delta_T}

\bibliographystyle{JHEP} 
\bibliography{main}

\end{document}